# Selective oxidation-induced strengthening of Zr/Nb nanoscale multilayers

M. A. Monclús[1,*], M. Callisti[2], T. Polcar[3,4], L.W. Yang[1], J. LLorca[1,5], and J. M. Molina-Aldareguía[1,*]

[1] IMDEA Materials Institute, c/Eric Kandel 2, 28906 Getafe, Madrid, Spain.
[2] Engineering Materials, Faculty of Engineering and the Environment, University of Southampton, Southampton SO17 1BJ, UK.
[3] National Centre for Advanced Tribology (nCATS), Faculty of Engineering and the Environment, University of Southampton, Southampton SO17 1BJ, UK
[4] Department of Control Engineering, Faculty of Electrical Engineering, Czech Technical University in Prague, Technická 2, Prague 6, Czech Republic.
[5] Department of Materials Science, Polytechnic University of Madrid. E.T.S. de Ingenieros de Caminos, 28040 Madrid, Spain.

* Corresponding authors.

## Abstract

The paper presents a new approach, based on controlled oxidation of nanoscale metallic multilayers, to produce strong and hard oxide/metal nanocomposite coatings with high strength and good thermal stability. The approach is demonstrated by performing long term annealing on sputtered Zr/Nb nanoscale metallic multilayers and investigating the evolution of their microstructure and mechanical properties by combining analytical transmission electron microscopy, nano-mechanical tests and finite element models. As-deposited multilayers were annealed at 350ºC in air for times ranging between 1 – 336 hours. The elastic modulus increased by ~ 20% and the hardness by ~ 42% after 15 hours of annealing. Longer annealing times did not lead to changes in hardness, although the elastic modulus increased up to 35% after 336 hrs. The hcp Zr layers were rapidly transformed into monoclinic $ZrO_2$ (in the first 15 hours), while the Nb layers were progressively oxidised, from top surface down towards the substrate, to form an amorphous oxide phase at a much lower rate. The sequential oxidation of Zr and Nb layers was key for the oxidation to take place without rupture of the multi-layered structure and without coating spallation, as the plastic deformation of the metallic Nb layers allowed for the partial relieve of the residual stresses developed as a result of the volumetric expansion of the Zr layers upon oxidation. Moreover, the development of residual stresses induced further changes in mechanical properties in relation to the annealing time, as revealed by finite element simulations.

1. Introduction

Nanoscale metallic multilayers have been studied in detail in the last 10 years due to their outstanding mechanical properties [1–4], especially for individual layer thickness below ≈100 nm, that originate from the high density of interfaces, that block dislocation transmission. From all possible metal combinations, fcc/bcc systems with incoherent interfaces, like Cu/Nb, have been widely studied, as they can achieve high strength at ambient temperature [1,5]. Other combinations, like the substitution of Cu by Zr, to produce Zr/Nb multilayers, are also promising as the Zr/Nb system is highly immiscible and the Zr/Nb interfaces are also expected to block dislocation transmission. The deformation behaviour of Zr/Nb multilayers fabricated by magnetron sputtering with different layer thicknesses was recently reported [6] and a few other studies on Zr/Nb multilayers are available. They are mainly focused on the hcp to bcc transformation of the Zr layers with decreasing bilayer thickness [7] or on their superconducting properties [8].

From the viewpoint of engineering applications, metallic multilayers are limited by thermal stability, especially in oxidizing environments. So far, the research on this area has been focussed on Cu/Nb nanoscale multilayers manufactured by magnetron sputtering or accumulated roll bonding. Mechanical tests in vacuum or Ar atmosphere have shown very good strength retention at high temperature [9,10] while the layer thickness and the ambient temperature strength were not modified by annealing up to 500 ºC in vacuum or inert atmospheres [11–14]. However, under oxidizing conditions, Cu/Nb multilayers are prone to degradation at temperatures as low as 300 ºC.

In this work, we study the influence of long term annealing in air on the microstructure and mechanical properties of Zr/Nb nanoscale metallic multilayers. Oxidation of Zr and Nb have been studied for many years, and it is well known that their oxide layers exhibit strong compressive residual stresses [15,16] as a consequence of the dramatic volume expansion of 56% for Zr and 154% for Nb, associated with their oxide formation [17], that lead to the spallation of oxide scales [18,19]. In this work, we show that in the case of Zr/Nb nanoscale multilayers, the oxidation of the Zr, owing to its more negative oxidation potential, occurs at a larger rate than that of Nb, and that the multilayered structure allows for an accommodation of the residual stresses developed during oxidation, preventing the spallation of the coatings. These results present a new approach, based on controlled oxidation of nanoscale metallic multilayers, to produce strong and hard oxide/metal nanocomposite coatings with high strength and good thermal stability.

2. Materials and experimental methods

The Zr/Nb multilayer was deposited on single crystal (100) Si wafers using a balanced magnetron sputtering apparatus (Kurt J. Lesker Company, Pennsylvania, US), with a total layer thickness of ≈ 1.35 μm. Monolithic Zr and Nb films were also deposited on the same substrate with thicknesses of 0.85 µm and 1.24 µm, respectively. Further details about the deposition process are reported elsewhere [6]. The multilayer and monolithic films were thermally annealed in air inside a muffle furnace at 350 ºC for annealing times $t_a$ = 2, 15, 48, 168 and 336 h. The as-deposited and annealed Zr/Nb multilayers presented a roughness, $R_a$, measured by AFM, of ≈ 3 and 5 nm, respectively.

The microstructure was evaluated by grazing incidence X-ray diffraction (XRD) and transmission electron microscopy (TEM). The XRD grazing angle ω was set at 1.5°, 3° and 5° in order to detect microstructural features at different depths. Diffraction data were collected by using a Rigaku SmartLab diffraction system (Rigaku Corporation, Japan) with Cu Kα radiation. The transmission electron microscope (JEOL JEM 2100) was operated at an accelerating voltage of 200 kV. TEM samples were prepared by using a focused ion beam (FIB) system (FEI Helios 600i). Scanning transmission electron microscopy high angle annular dark field (STEM-HAADF) images of the cross-sections were obtained by using the STEM detector in the FIB system with the electron gun operated at 30 kV. A JEOL ARM200F (cold-FEG) TEM/STEM operated at 200 kV and equipped with a Gatan GIF spectrometer and a 100 mm$^2$ Centurion EDX detector (Thermo Fisher Scientific Inc., Madison, Wisconsin, USA) was also used for STEM imaging and chemical analyses. EELS acquisitions were performed with an energy dispersion of 0.05 eV/channel for map acquisition in the low loss energy region, while 0.25 eV/channel was used to acquire data for line-scan in the low and high loss energy range.

Nanoindentation measurements were performed using the Nanotest Platform 3 Instrument (Micromaterials, Wrexham, UK) equipped with a diamond Berkovich indenter. Reported hardness and elastic modulus are an average of ten indents performed by using a maximum load of 5 mN (for the Zr/Nb multilayer) and 2 mN (for monolithic Zr and Nb coatings) with a loading, holding and unloading times of 20, 10 and 5 s, respectively. The maximum load was chosen in order to keep $h_c/t < 0.1$, where $h_c$ is the contact depth and $t$ the film thickness. The reported elastic modulus of the layers was corrected to eliminate substrate effects by using the Hay & Crawford model [20].

## 3. Results and discussion

### 3.1. Microstructure evolution upon annealing in air

A STEM-HAADF cross-sectional image of the as-deposited Zr/Nb multilayer along with the corresponding selected area diffraction (SAD) pattern are presented in Fig. 1. Nb appears brighter in the STEM-HAADF micrographs owing to its higher density compared to Zr. The layered structure is well defined with flat interfaces close to the Si substrate. Layer waviness increased as deposition progressed due to shadowing effects inherent to the sputtering processes. The layer thicknesses were $t_{Nb} \sim 30$ nm and $t_{Zr} \sim 45$ nm. The SAD pattern (indexed as hcp Zr and bcc Nb) indicates that Zr and Nb layers were nanocrystalline. Zr nanograins presented two main preferred orientations, i.e. Zr{10-10} and Zr{0002} parallel to the substrate while Nb nanograins were randomly oriented. The high resolution TEM micrograph in Fig. 1b, which shows the first Zr/Nb bilayer, revealed that the multilayer exhibited a polycrystalline columnar growth with irregular lateral grain sizes ranging between 10 – 30 nm.

Oxidation of the sputtered monolithic Zr coatings resulted in spallation of large areas of the coatings just after 2 h of exposure to air at 350 ºC. In those areas that survived after spallation, it was possible to determine that the coatings rapidly transformed into monoclinic $ZrO_2$, resulting in a thickness increase of 24% and 32% after 2 h and 48 h annealing at 350 ºC, respectively. The large thickness increase was a clear manifestation of the large volumetric expansion associated with the formation of the oxide, which is expected to be 56% for Zr [16], leading to the development of large compressive stresses responsible for the spallation of the coatings, as it is usually observed for monolithic Zr [18].

On the other hand, the Zr/Nb multilayers did not suffer any spallation or fracture, and survived the annealing process in air at 350ºC, for times as long as 336 h. Moreover, detailed TEM studies revealed that the oxidation process underwent without loss of the multilayer structure. STEM-HAADF cross-sectional images and EDX line scan analyses of the as-deposited and annealed Zr/Nb multilayers are shown in Fig. 2. The total thickness of the multilayer increased with annealing time ($t_a$), and this phenomenon was mainly associated with a thickness increase of Zr layers from ~ 45 nm for the as-deposited multilayer to ~ 50 nm and ~ 55 nm for $t_a$ = 15 h and 168 h, respectively. EDX line scans showed the preferential oxidation of Zr layers, while the Nb oxidised at a much slower rate. Only the first 2 layers ($t_a$ =15 h), 3 layers ($t_a$ =48 h) and 5 layers ($t_a$ =168 h) of Nb experienced a thickness increase from ~ 30 to ~ 50 nm due to oxidation. The remaining Nb layers showed negligible thickness variation during annealing at 350ºC, although they show a substantial uptake of oxygen, as seen in the EDX profiles, even below the fully oxidized region.

Fig. 3 shows a bright field cross-sectional TEM image of the Zr/Nb multilayer annealed for 168 h. The area illustrated corresponds to a region about 450 – 650 nm below the surface where there is a clear boundary between the fully oxidised and partially oxidised Nb layers. The $ZrO_2$ layers appear to maintain the polycrystalline structure throughout the multilayer thickness, whereas the Nb layers adopt an amorphous structure after full oxidation. This is corroborated by the SAD patterns for the two regions: above the boundary (near the surface), the diffraction rings are formed by a mixture of continuous rings and discrete spots with all but one indexed as the monoclinic $ZrO_2$ phase. $ZrO_2$ may exist at room temperature in two allotropic forms: (1) tetragonal, which can only be stabilised by high compressive stresses near the metal/oxide interface [21] and (2) monoclinic, which is the most stable crystal structure of the zirconia phase at room temperature [22] and the dominant phase in sputtered $ZrO_2$ films [23]. There is one faint ring that could be attributed to few remaining Zr (110) crystals. The information in Fig. 3 clearly shows that oxidized Nb layers are thicker and exhibit an amorphous structure in the region near the surface, where no Nb rings can be found in the DP. The evolution of the Nb layers upon annealing follow the expected oxidation sequence for this metal. First, the metallic Nb lattice undergoes a substantial oxygen uptake [19], as demonstrated by the EDX profiles of Fig. 2. These Nb layers suffer a negligible lattice distortion upon oxygen uptake, as shown in the DP of Fig. 3 below the dashed line, where the Nb rings can be identified in the SAD pattern, and the diffraction rings appear more continuous. Hence, these layers are labelled as $NbO_x$ to reflect their high oxygen content. Above a certain critical oxygen content, full oxidation of the Nb layers takes place, presumably by the formation of amorphous $Nb_2O_5$ [24], resulting in a large volumetric expansion, as can ben seen in the fully oxidized region near the surface of Fig. 3. The formation of amorphous $Nb_2O_5$ resulted in a large increase in thickness, from 30 nm to 50 nm, even bigger than for the Zr layers, as expected from the theoretical volumetric expansion of 154% for oxidation of Nb, as opposed to 56% for Zr.

In order to shed light about the oxidation process, the Zr/Nb multilayer annealed for 168 h at 350ºC was investigated by EELS. Low- and high-loss electron energy loss spectra were collected across the boundary between the totally and partially oxidised regions. Fig. 4 shows the EEL spectra extracted from the Zr and Nb layers above and below this boundary.

EEL spectra acquired on the Zr layers above the boundary (spectrum 4 in Fig. 4) exhibit very similar spectral features, which are in good agreement with the EEL spectrum reported for $ZrO_2$ [25]. The presence of overlapping peaks located at $\approx$ 15 eV and at $\approx$ 25 eV was attributed to the splitting of energy levels within conduction/valance bands of the oxide [25] reflecting the non-metallic nature of the metal-oxygen bonds. The EEL spectra for Zr layers located further below the boundary (spectrum 3 in Fig. 4) were very similar to those above the boundary (spectrum 4 in Fig. 4). A different scenario was found for Nb layers, where the broad Nb $N_{2,3}$ edge at $\approx$ 43 eV observed in spectrum 1 (Fig. 4) shifted towards higher energy losses in spectrum 2 (Fig. 4). This shift can be due to the presence of oxygen [25]. Furthermore, the presence of extra peaks beside the plasmon peak in spectrum 2 suggests a change in bonding from metal-metal to metal-oxygen. These results indicate that metallic Nb may form a sub-oxide first, while an oxide phase forms when enough oxygen is accommodated in the Nb layers, as .

An EELS linescan across the whole multilayer is shown in Fig. 5a. Although absolute quantification is not precise due to the overlapping of spectral features, the oxygen line profile clearly indicates that most of the oxygen is stored in the Nb layers in the heavily oxidised region above the boundary. Quantification of EEL spectra extracted from spectrum images on Nb layers above the boundary confirmed a $Nb_2O_5$ composition in the fully oxidized region, while the oxygen content in the Nb layers decayed with distance from the surface. On the other hand, Zr layers accommodated more oxygen than Nb layers below the boundary, but the oxygen content of the Zr layers was found to be approximately constant above and below the boundary. The annealing process did not cause any relevant mixing between constituent elements as can be observed in the EDX elemental maps in Fig. 5b. They show a layer thickness increase in the heavily oxidised region, especially for Nb layers, which absorbed a larger amount of oxygen in this region. On the other hand, most of the oxygen was stored in the Zr layers below the boundary. These results corroborated the preferential oxidation of Zr with respect to Nb, which instead undergoes a much slower oxidation process.

The TEM observations were corroborated by the XRD patterns of the as-deposited and annealed Zr/Nb multilayers shown in Fig. 6. Only hcp Zr and bcc Nb were detected in the as-deposited Zr/Nb with well-defined Zr(002) and Nb(110) peaks, in agreement with the SAD patterns. After annealing for 15 h, the monoclinic $ZrO_2$ phase became dominant at the expense of the Zr phase, while the intensity of Nb peaks decreased but to a much lesser extent. This indicates that Nb layers did not appear to be significantly oxidized after annealing for 15 h at 350ºC, while all the Zr layers were mostly transformed to $ZrO_2$ in agreement with the TEM observations. After 168 h, the Nb peaks also disappeared, indicating that the Nb layers near the surface have been oxidized and the $ZrO_2$ peaks dominate the XRD patterns with little traces of Zr and Nb metallic phases.

In-plane residual stresses in the annealed Zr/Nb multilayers were estimated qualitatively from the shifts of the peaks observed in the XRD θ-2θ patterns (not shown here) by comparing the position of the measured patterns with respect to the stress-free reference standards (ICDD PDF-2 database). In the as-deposited condition ($t_a = 0$ h), the Zr and Nb peaks shifted towards lower angles indicating that the Zr and Nb were subjected to intrinsic compressive residual stress that developed as a result of the growing process, as thermal stresses should be negligible due to the relatively low deposition temperature. Upon annealing, the Nb peaks shifted towards higher 2θ angles, thus indicating that the Nb layers developed in-plane tensile stresses. Considering the large volumetric expansion associated with the formation of $ZrO_2$ (the layer thickness increased from ~ 45 to ~ 55 nm), it is surprising that the formation of $ZrO_2$ did not result on spallation of the coatings, as it was the case for the oxidation of the monolithic Zr coatings [18]. The fact that no delamination between layers was observed even after annealing for 2 weeks far from the multilayer surface, indicated that the volumetric expansion was partially accommodated by the plastic deformation of the metallic Nb layers. This would lead to an increase in the compressive stresses in the Zr layers and to the development of tensile stresses in Nb, in agreement with the qualitative information about the residual stresses obtained from the XRD θ-2θ patterns. Furthermore, the state of residual stresses left after the oxidation of the Zr layers might favour the accommodation of the large stresses that are expected to develop when the Nb layers start to oxidize. The fact that the Nb and $ZrO_2$ layers are, after oxidation of the Zr layers, left under tensile and compressive residual stresses, respectively, might help accommodate the large compressive stresses expected upon oxidation of the Nb layers (that undergo a large increase in thickness, from ~ 30 to ~ 50 nm), and the corresponding tensile stresses expected in the $ZrO_2$ layers. However, at this stage, the $ZrO_2$ layers are not expected to accommodate the volumetric expansion of the oxidizing Nb layers by plastic deformation. As a matter of fact, it is interesting to note that, upon annealing, defects in the form of cracks and delaminations appeared in the fully oxidised region near the surface and, in particular, in the $ZrO_2$ layers (Fig. 7a). Some defects exhibited a random shape and distribution within the $ZrO_2$ layers, while others were located along to the $Nb_2O_5/ZrO_2$ interfaces and presented an elongated shape. Some of the cracks formed along interfaces tended to coalesce, although in most cases isolated cracks were observed (Fig. 7b and 7c). Hence, these cracks probably developed as a result of the large volumetric expansion associated with the formation of $Nb_2O_5$ near the surface, as they are not present far from the surface where only the Zr layers were oxidized and the Nb layers remained metallic.

*3.2. Hardness and elastic modulus evolution upon annealing in air*

The ambient temperature hardness and elastic modulus of the as-deposited and annealed Zr/Nb multilayers and monolithic Zr and Nb films are shown in Figs. 8a and 8b, respectively, as a function of annealing time ($t_a$). The layered structure provided an additional strengthening ($H_{Zr/Nb}$ = 7.7 GPa) with respect to the hardness given by the rule-of-mixtures from those of the Zr and Nb constituents ($H_{Zr}$ = 6.1 GPa and $H_{Nb}$ = 6.7 GPa) in the as-deposited condition. A similar strengthening effect was observed on other multilayers with incoherent interfaces, because interfaces stand as a major obstacle to dislocation glide due to the large elastic and lattice mismatch between Zr and Nb [26]. Annealing of the Zr and Nb films as well as of the Zr/Nb multilayer led to a large increase in hardness (Fig. 8a). The largest change was found in the monolithic Zr layer, whose hardness increased from 6.1 GPa to 12.2 GPa after 2h of annealing at 350ºC, following the Zr to $ZrO_2$ oxidation. However, adhesion problems of monolithic Zr layer to the Si substrate resulted in the delamination of the coating just after a few hours of annealing. The monolithic Nb layer experienced a moderate increase in hardness after oxidation. Delamination of the Nb layer from the Si substrate did not take place even after $t_a$ = 336 h., likely due to the much slower oxidation rate of Nb, which allowed the accommodation of the compressive stresses induced by the volume increase due to oxidation. Finally, the hardness of the Zr/Nb multilayer was also increased by annealing at 350ºC and reached a peak of 10.9 GPa for $t_a$ = 15 h. Longer annealing times did not change the hardness and the multilayer remained well bonded to the Si substrate.

The elastic modulus of the Zr/Nb multilayer ($E_{Zr/Nb}$ = 147.7 GPa) was higher than that calculated by the rule-of-mixture from those of the Zr and Nb constituents ($E_{Zr}$ = 135.0 GPa and $E_{Nb}$ = 113.0 GPa), Fig. 8b. Annealing at 350ºC led to an increase in the elastic modulus of the Zr and Nb monolithic films as well as of the Zr/Nb multilayer. The elastic modulus of Zr increased rapidly with annealing time due to the rapid oxidation of Zr while the increase in the modulus of Nb took longer due to the slower oxidation process. The elastic modulus of the annealed Zr/Nb multilayer exhibited similar trends to those found for the monolithic Nb, except for a more abrupt increase for $t_a$ > 100 h.

## 4. Numerical model and discussion

It is clear from the above results that high temperature oxidation led to a large strengethening of the Zr/Nb multilayer due to the enhancement of mechanical properties of individual layers after oxidation and to the development of internal stresses in the layers. In order to clarify the role played by these mechanisms, finite element simulations were performed to quantify the effect of the number of oxide layers and of the internal stresses on the strength of the annealed Zr/Nb multilayers.

*4.1. Geometrical model*

Numerical simulations of the nanoindentation tests of the Zr/Nb multilayer were perfomed using Abaqus [27]. The schematic of the Zr/Nb multilayer model is depicted in Fig. 9. The 2D axisymmetric model included a rigid conical indenter (semi-angle of 70.3º), a Si substrate and 18 layers of Zr ($ZrO_2$) and Nb ($Nb_2O_5$). The width of the geometrical model was large enough to eliminate boundary effects. Numerical simulations of indentation were carried out assuming that the Zr layers were progressively oxidized from the top surface. Since the oxidization of Nb took place at a much slower rate, as compared to Zr, it was assumed that the first Nb layer was fully oxidised only after all Zr layers were oxidized. Thus, hardness values were computed for the as-deposited Zr/Nb multilayer, $ZrO_2$-Zr/Nb multilayers and $ZrO_2$/ $Nb_2O_5$-Nb multilayers. The number of Zr and Nb oxidised layers chosen for the simulations was 3, 6, 9, 12, 15 and 18.

The thickness of the Zr and Nb layers was 45 nm and 30 nm, respectively, and the thickness of the corresponding oxidized layers was 55 nm, based on TEM observations. The model was discretized with 14418 4-node bilinear axisymmetric quadrilateral elements with reduced integration (CAX4R). The interfaces between layers were assumed to be perfect, and interface sliding or fracture was not included in the model.

Zr, Nb and their oxides were assumed to behave as isotropic, elasto-perfectly plastic solids. The elastic modulus was obtained from nanoindentation results of the corresponding monolithic layers. The yield stress was estimated from the nanoindentation hardness assuming a Tabor factor of 2.9 [28], as listed in Table I. The Poisson's ratio of Zr($ZrO_2$) and Nb($Nb_2O_5$) were 0.4 and 0.34, respectively. The Si wafer was assumed to behave as an isotropic elastic solid, with elastic modulus of 187 GPa and Poission's ratio of 0.18.

The simulation of the indentation process was carried out by moving vertically the rigid conical indentor. The penetration depth was set to 140 nm, which corresponds to ~ 10% of the total multilayer thickness, in order to eliminate substrate effects. The bottom boundary was completely fixed in space while the lateral boundary was unconstrained during indentation. The first indentation was carried out in the metallic Zr/Nb multilayer.

The second indentation was carried out assuming that the first three layers of Zr have transformed into $ZrO_2$. To this end, the mechanical properties of the first three layers were modified and, in addition, it was assumed that these three layers have experienced a volume increase due to oxidation of $\Delta V_{ZrO2} = 66.7\%$, which would lead to experimental linear increase in the layer thickness from 45 nm to 55 nm ($\Delta V_{ZrO2}/3$). This procedure was repeated by including three layers of oxidised Zr in the model until all the Zr layers were oxidised. Afterwards, the same strategy was followed with the Nb layers, starting from the top surface. The volume increase in the case was $\Delta V_{Nb2O5} = 250\%$, as the thickness of the Nb layers during oxidation increased from 30 nm to 55 nm ($\Delta V_{Nb2O5}/3$). The Oliver & Pharr method was used to extract multilayer hardness from the simulated force-displacement curves [29]. Finally, in order to ensure the robustness of the simulations results, the model sensitivity to potential variations in layer thickness was assessed by varying them within ±10%. The simulations results were found to be robust to such potential errors.

*4.2. Numerical results*

The simulated nanoindentation curves are depicted in Fig. 10 for the as-deposited Zr/Nb multilayers, the $ZrO_2$-Zr/Nb multilayers with 6 $ZrO_2$ layers and the $ZrO_2$/Nb-$Nb_2O_5$ multilayers with 6 $Nb_2O_5$ layers. For each oxidised material, two simulations were presented in which the internal stresses associated with the increase in volume due to oxidation were and were not taken into account. In the latter case, the only difference introduced by oxidation was the change in the mechanical properties of the oxidized layers but the internal stresses induced by the volume increase were not included in the model.

The numerical simulations in Fig. 10 show that the multilayer deformation was very sensitive to both the layer oxidation and to the presence of internal stresses induced by the volume increase associated with oxidation. In fact, the increment in maximum load upon nanoindentation in the $ZrO_2$-Zr/Nb multilayer with 6 $ZrO_2$ layers was mainly due to the oxidation of Zr to $ZrO_2$ while the internal stresses played a noticeable but secondary role. In the case of the $ZrO_2$/ $Nb_2O_5$-Nb multilayers with 6 $Nb_2O_5$ layers, the contribution of the internal stresses to the maximum load was as high as that of the oxidation of the Nb to $Nb_2O_5$. The contribution of oxidation to the hardness was clearly due to the higher stiffness and strength of the oxidised layers, while the influence of the residual stresses can be understood from the contour plots of the radial stress plotted in Figs. 12a and 12b for the unloaded $ZrO_2$/Nb multilayer and the $ZrO_2$/ $Nb_2O_5$-Nb with 6 $Nb_2O_5$ layers, respectively. Neglecting the compressive residual stresses introduced in the Zr and Nb layers during deposition, oxidation of the Zr layer to $ZrO_2$ during high temperature annealing led to very high compressive stresses in the $ZrO_2$ layers ($\approx$ 4.2 GPa) and tensile stresses in the Nb layers ($\approx$ 0.5 GPa) (Fig. 12a), in agreement with the qualitative information in the XRD θ-2θ spectra. Oxidation of the first six layers of Nb to $Nb_2O_5$ reduced the compressive stresses in the $ZrO_2$ layers and introduced high compressive stresses in the upper $Nb_2O_5$ layers ($\approx$ 3.1 GPa) (Fig. 12b). Thus, all the layers near the surface in the $ZrO_2$/ $Nb_2O_5$-Nb were subjected to high compressive internal stresses, which were opposed to the penetration of the nanoindenter and increased the hardness of the multilayer (Fig. 12c). In fact, some degree of pile-up was observed at the free edge of the nanoindentation imprint due to the large compressive stresses in the oxidised layers.

The numerical and experimental variation of the multilayer hardness as a function of the number of oxidised Zr and Nb layers is plotted in Fig. 12. The simulation results with and without including the internal stresses induced by the volume change due to oxidation showed a two-step increase in hardness with the number of oxidised layers. The hardness grew initially with the number of $ZrO_2$ layers due to the presence of the harder $ZrO_2$ layers until a plateau was reached after ~ 10 $ZrO_2$ layers were formed. The plateau in hardness after 10 $ZrO_2$ layers can be explained by the limited penetration depth of the nanoindentor. The stress field is mainly controlled by the layers around the indenter tip and the properties of the layers far away from the deformed zone around the tip have little effect on the measured hardness. A steeper growth in the multilayer hardness was observed once the first three Nb layers were oxidized and this increase was also attributed to the harder $Nb_2O_5$ layers, until a plateau was reached once 6 $Nb_2O_5$ layers were formed. It is evident from Fig. 12 that the stresses associated with volume expansion during oxidation have a much larger effect on the multilayer hardness after the Nb layers began to oxidise. Note too that the hardness obtained from the simulations (considering the stresses associated with layer expansion) correlates very well with the experimental results while the Zr layers oxidise, but it is significantly higher than the experimental results after Nb layers start to oxidise. These differences can be attributed to the development of damage in the form of cracks and holes in the multilayers during the oxidation of Nb to $Nb_2O_5$ (Fig. 5), which can release the internal stresses. It should be noticed that plastic deformation of Nb could absorb the volume change during the initial oxidation of the Zr to $ZrO_2$. However, it was more difficult for $ZrO_2$ to absorb the volume increased associated with the oxidation of Nb, and this led to the development of cracks along the interface between layers. In any case, damage is not included in the model, which always assumes a perfectly bonded interface, the release in internal stresses due to cracking was not accounted for in the simulations, leading to the overestimation of the hardness once Nb layers start to oxidise.

## 5. Conclusions

A Zr/Nb nanoscale metallic multilayer with a hcp/bcc nanocrystalline structure was synthesised by magnetron sputtering, with layers thicknesses of 45/30 nm. The as-deposited multilayer exhibited a hardness and elastic modulus above those of the constituent elements. Zr layers transformed from an hexagonal closed-packed structured to a monoclinic $ZrO_2$ phase after just a few hours of annealing in air at 350ºC, while the Nb layers oxidised at a much slower rate. Oxidised multilayers were thicker and the hardness and the elastic modulus of the annealed Zr/Nb multilayer increased by near 40% after long term annealing. The oxidised multilayer exhibited good thermal stability without apparent deterioration and maintained the high hardness and modulus even after annealing for 336 h at 350ºC.

Experimental observations showed that the all the Zr layers were rapidly transformed into monoclinic $ZrO_2$ (in the first 15 hours), while the progressive oxidation of the Nb layers from the top surface to the bottom took much longer. The increase in hardness with the annealing time was studied by means of finite element models, which take into account change in mechanical properties of the layers due to oxidation (Zr was transformed to $ZrO_2$ and Nb to $Nb_2O_5$) and the development of internal stresses to accommodate the volume increase associated with the sequential oxidation of Zr and Nb layers during annealing. It was found that both mechanisms contributed to the observed increase in hardness during short term annealing although the first one (oxidation of Zr to $ZrO_2$) was dominant. During long term annealing, the finite element models showed that further strengthening was possible due to the synergistic contributions of the transformation of Nb to $Nb_2O_5$ and the development of compressive stresses in both $ZrO_2$ and $Nb_2O_5$ layers to accommodate the volumetric expansion of the Nb layers during oxidation. However, the experimental results did not show any further increase in hardness during long term annealing (beyond 15 h). This behaviour was associated to the development of damage at the interface between $ZrO_2$ and $Nb_2O_5$ layers, and the corresponding release of the internal stresses, as $ZrO_2$ was not able to accommodate by plastic deformation the volumetric expansion during the oxidation of Nb. In summary, the sequential oxidation of Zr and Nb layers was key for the oxidation to take place without rupture of the multi-layered structure and without coating spallation, as the plastic deformation of the metallic Nb layers allowed for the partial relieve of the residual stresses developed as a result of the volumetric expansion of the Zr layers upon oxidation. These results introduce a new approach, based on controlled oxidation of nanoscale metallic multilayers, to obtain ceramics/metal nanocomposites with enhanced thermal stability and mechanical properties.

**Acknowledgements**

The financial support from the European Union through the RADINTERFACES project (Grant agreement No. 263273) and from the Spanish Ministry of Economy and Competitiveness (MAT2012-31889) are gratefully acknowledged. In addition, the support from the European Research Council (ERC) under the European Union Horizon 2020 research and innovation programme (Advanced Grant VIRMETAL, grant agreement No. 669141) in the last stages of this investigation is also acknowledged. The South of England Analytical Electron Microscope (EPSRC Grant code EP/K040375/1) is acknowledged for access to the EM facilities.

**References**

[1] A. Misra, J.P. Hirth, R.G. Hoagland, Length-scale-dependent deformation mechanisms in incoherent metallic multilayered composites, Acta Mater. 53 (2005) 4817–4824.

[2] Y.Y. Lu, R. Kotoka, J.P. Ligda, B.B. Cao, S.N. Yarmolenko, B.E. Schuster, The microstructure and mechanical behavior of Mg/Ti multilayers as a function of individual layer thickness, Acta Mater. 63 (2014) 216–231.

[3] J.Y. Zhang, G. Liu, J. Sun, Strain rate sensitivity of nanolayered Cu/X (X=Cr,Zr) micropillars: Heterophase interfaces/nanotwin boundaries effects, Mat. Sci. Eng. A. 612 (2014) 28–40.

[4] Y. Chen, Y. Liu, C. Sun, K.Y. Yu, M. Song, H. Wang, Microstructure and strengthening mechanisms in Cu/Fe multilayers, Acta Mater. 60 (2012) 6312–6321.

[5] N. A. Mara, D. Bhattacharyya, P. Dickerson, R.G. Hoagland, A. Misra, Deformability of ultrahigh strength 5 nm Cu/Nb nanolayered composites, Appl. Phys. Lett. 92 (2008) 231901.

[6] E. Frutos, M. Callisti, M. Karlik, T. Polcar, A Length-scale-dependent mechanical behaviour of Zr/Nb multilayers as a function of individual layer thickness, Mat. Sci. Eng. 632 (2015) 137–146.

[7] G.B. Thompson, R. Banerjee, S. A. Dregia, H.L. Fraser, Phase stability of bcc Zr in Nb/Zr thin film multilayers, Acta Mater. 51 (2003) 5285–5294.

[8] R. Banerjee, P. Vasa, G.B. Thompson, H.L. Fraser, P. Ayyub, Proximity effect in Nb/Zr multilayers with variable Nb/Zr ratio, Solid State Commun. 127 (2003) 349–353.

[9] N.A. Mara, T. Tamayo, A. V. Sergueeva, X. Zhang, A. Misra, A.K. Mukherjee, The effects of decreasing layer thickness on the high temperature mechanical behavior of Cu/Nb nanoscale multilayers, Thin Solid Films. 515 (2007) 3241–3245.

[10] M.A. Monclús, S.J. Zheng, J.R. Mayeur, I.J. Beyerlein, N. A. Mara, T. Polcar, J. Llorca, J.M. Molina-Aldareguía, Optimum high temperature strength of two-dimensional nanocomposites, APL Mater. 1 (2013) 052103.

[11] J.S. Carpenter, S.J. Zheng, R.F. Zhang, S.C. Vogel, I.J. Beyerlein, N. A. Mara, Thermal stability of Cu–Nb nanolamellar composites fabricated via accumulative roll bonding, Philos. Mag. 93 (2013) 718–735.

[12] S. Zheng, I.J. Beyerlein, J.S. Carpenter, K. Kang, J. Wang, W. Han, High-strength and thermally stable bulk nanolayered composites due to twin-induced interfaces, Nat. Commun. 4 (2013) 1696.

[13] A. Misra, R.G. Hoagland, H. Kung, Thermal stability of self-supported nanolayered Cu/Nb films, Philos. Mag. 84 (2004) 1021–1028.

[14] N. A. Mara, A. Misra, R.G. Hoagland, A. V. Sergueeva, T. Tamayo, P. Dickerson, High-temperature mechanical behavior/microstructure correlation of Cu/Nb nanoscale multilayers, Mater. Sci. Eng. A. 493 (2008) 274–282.

[15] L. Kurpaska, J. Favergeon, L. Lahoche, G. Moulin, M. El Marssi, J.M. Roelandt, Zirconia layer formed by high temperature oxidation of pure zirconium: Stress generated at the zirconium/zirconia interface, Oxid. Met. 79 (2013) 261–277.

[16] C. Roy, B. Burgess, A study of the stresses generated in zirconia films during the oxidation of zirconium alloys, Oxid. Met. 2 (1970) 235–261.

[17] R.E. Bedworth, N.B. Pilling, The Oxidation of Metals at High Temperatures, J. Inst. Met. 29 (1923) 529.

[18] M. Parise, O. Sicardy, G. Cailletaud, Modelling of the mechanical behavior of the metal–oxide system during Zr alloy oxidation, J. Nucl. Mater. 256 (1998) 35–46.

[19] L.J. Weirick, The effect of stress on the low-temperature oxidation of Niobium, Retrosp.

Theses Diss. Paper 3617 (1969).

[20] J. Hay, B. Crawford, Measuring substrate-independent modulus of thin films, J. Mater. Res. 26 (2011) 727–738.

[21] J.A. Pardo, R.I. Merino, V.M. Orera, J.I. Pena, C. Gonzalez, J.Y. Pastor, J. Llorca, Piezospectroscopic study of residual stresses in Al2O3-ZrO2 directionally solidified eutectics, J. Am. Ceram. Soc. 83 (2000) 2745–2752.

[22] J.Y. Park, J.K. Heo, Y.C. Kang, The properties of RF sputtered Zirconium Oxide thin films at different plasma gas ratio, Bull. Korean Chem. Soc. 31 (2010) 397–400.

[23] P. Gao, L.J. Meng, M.P. Dos Santos, V. Teixeira, M. Andritschky, Influence of sputtering pressure on the structure and properties of ZrO2 films prepared by rf reactive sputtering, Appl. Surf. Sci. 173 (2001) 84–90.

[24] S. Dub, V. Starikov, Elasticity module and hardness of niobium and tantalum anode oxide films, Funct. Mater. 14 (2007) 7–10.

[25] Y.P. Lin, O.T. Woo, Oxidation of Zr and related phases in Zr-Nb alloys: An electron microscopy investigation, J. Nucl. Mater. 277 (2000) 11–27.

[26] L.H. Friedman, Exponent for Hall–Petch behaviour of ultra-hard multilayers, Philos. Mag. 86 (2006) 1443–1481.

[27] M.P. Hibbit, Carlsson, Sorensen, Abaqu´s User Manual 2011, Abaqu´s User Man. (2011).

[28] D. Tabor, The Hardness of Metals, Oxford University Press (2000).

[29] W. Oliver, G. Pharr, An improved technique for determining hardness and elastic modulus using load and displacement-sensing indentation systems, J. Mater. Res. 7 (1992) 1564–1583.

**List of figures**

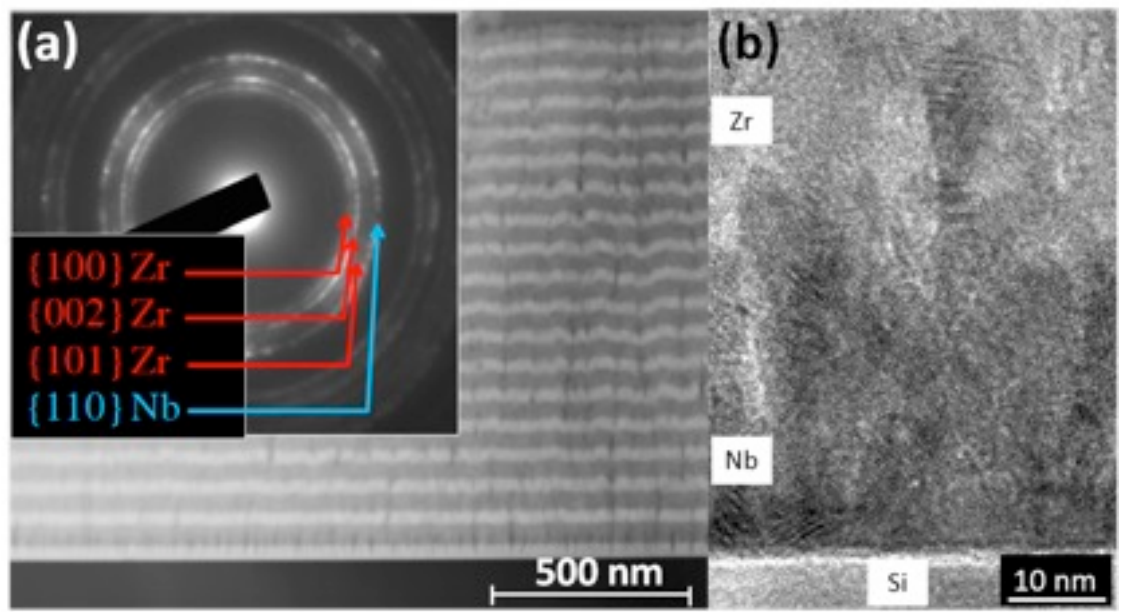

Fig. 1. (a) STEM-HAADF (FIB-STEM) micrograph of the Zr/Nb multilayer with the SAD pattern (inset). (b) High resolution TEM micrograph of the first Zr/Nb bilayer.

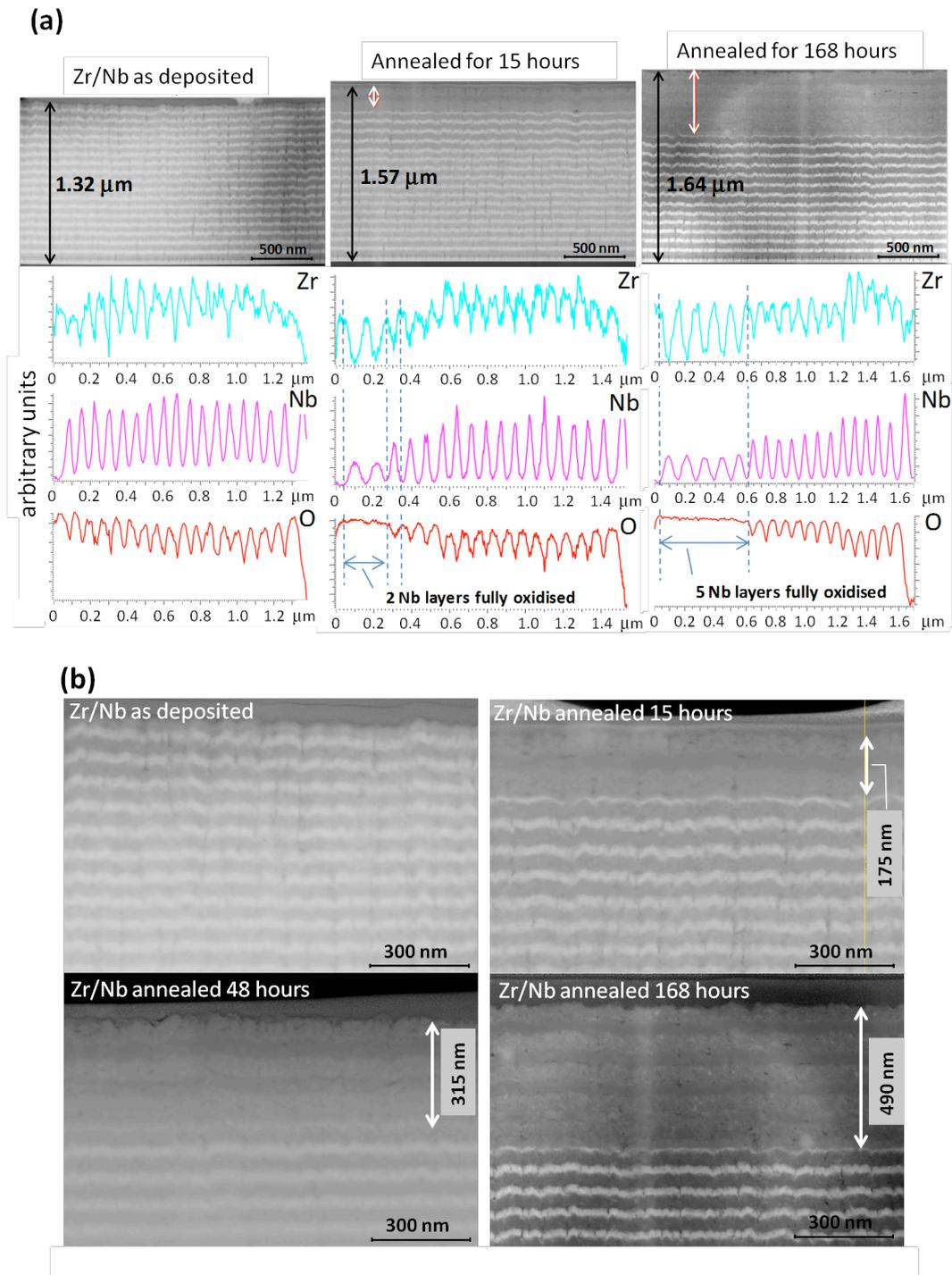

Fig. 2. (a) STEM-HAADF (FIB-STEM) images for the as-deposited and annealed Zr/Nb multilayers along with EDX line scans profiles of Zr, Nb and O in each layer. Nb appears brighter than Zr in the micrographs owing to its higher density. The blue arrows indicate the extent of the multilayer region where all the Zr and Nb layers are fully oxidised. (b) Higher magnification images of the STEM – HAADF micrographs. The white arrows indicate the region where Zr and Nb were fully oxidized.

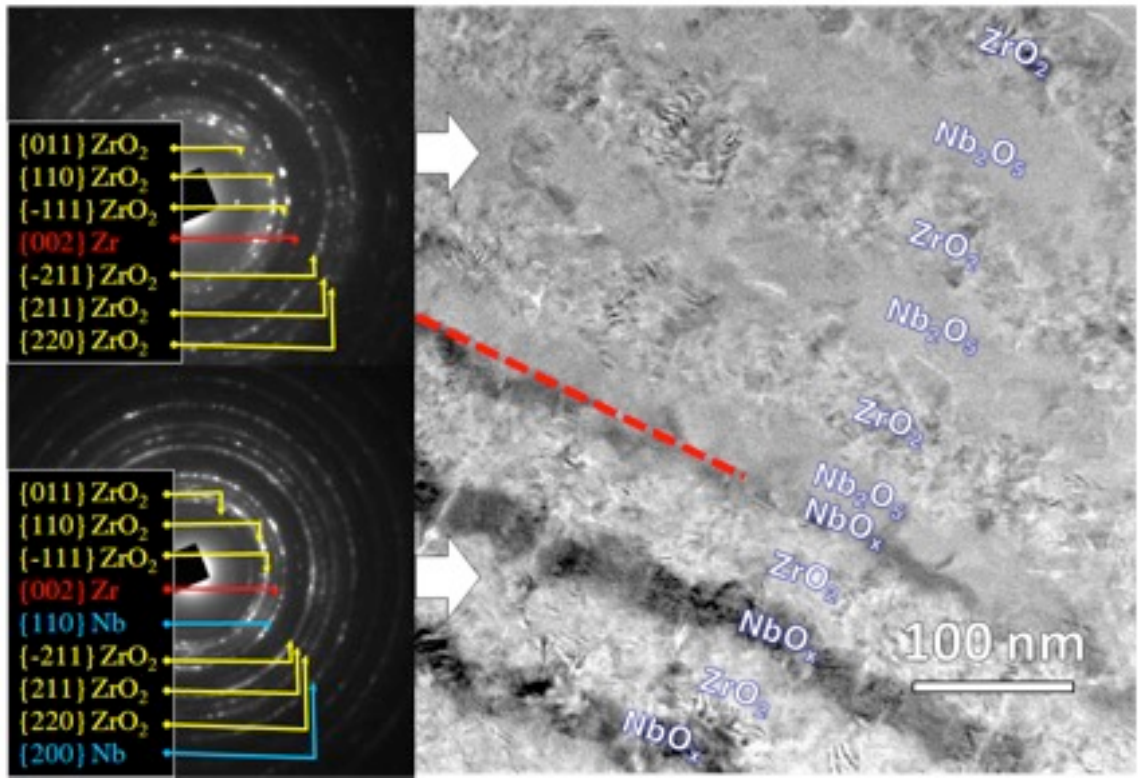

Fig. 3. TEM bright field cross sectional micrographs of the Zr/Nb multilayer annealed for 168 h at 350ºC. The discontinuous red line marks the boundary for the oxidation of Nb. SAD patterns correspond to the top region near the surface (Nb is fully oxidised) and to the bottom region (Nb is partially oxidised). The diameter of the aperture was 300 nm.

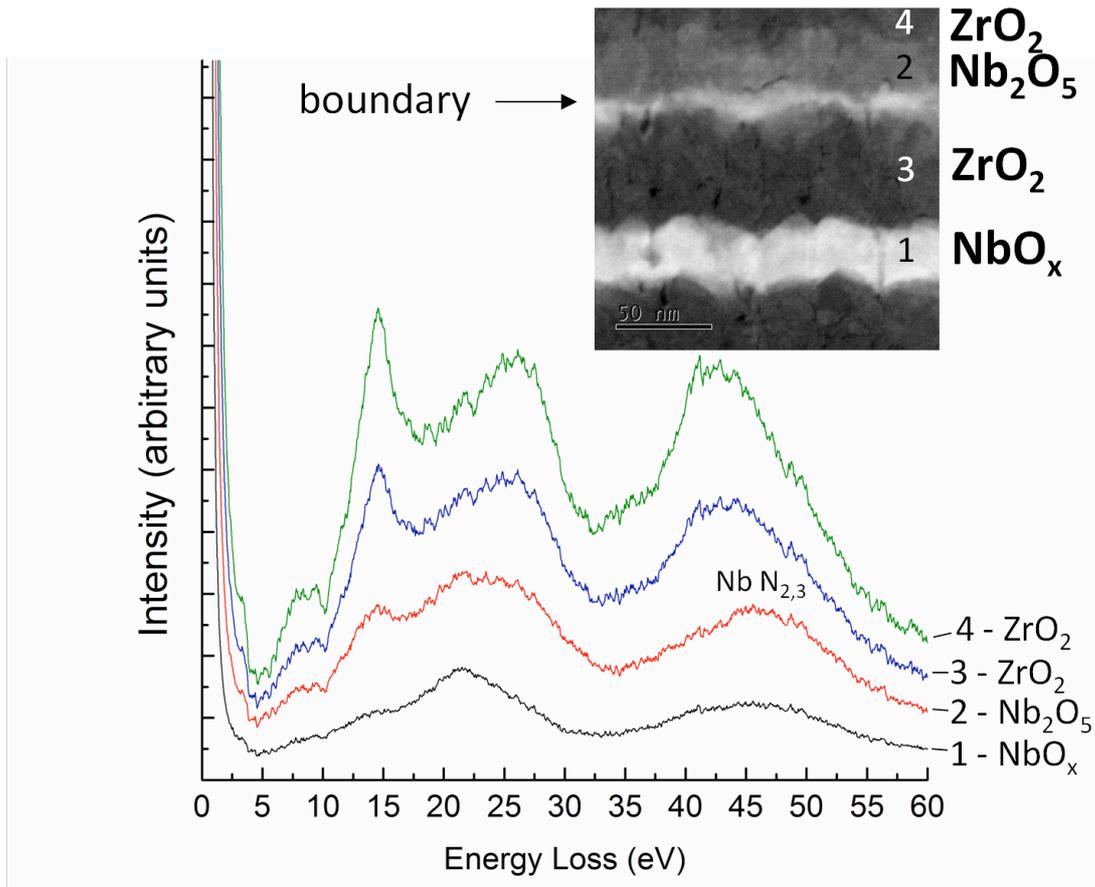

Fig. 4. Low-loss EEL spectra acquired on Zr and Nb layers above and below the boundary between heavily and non-heavily oxidised regions of the multilayer.

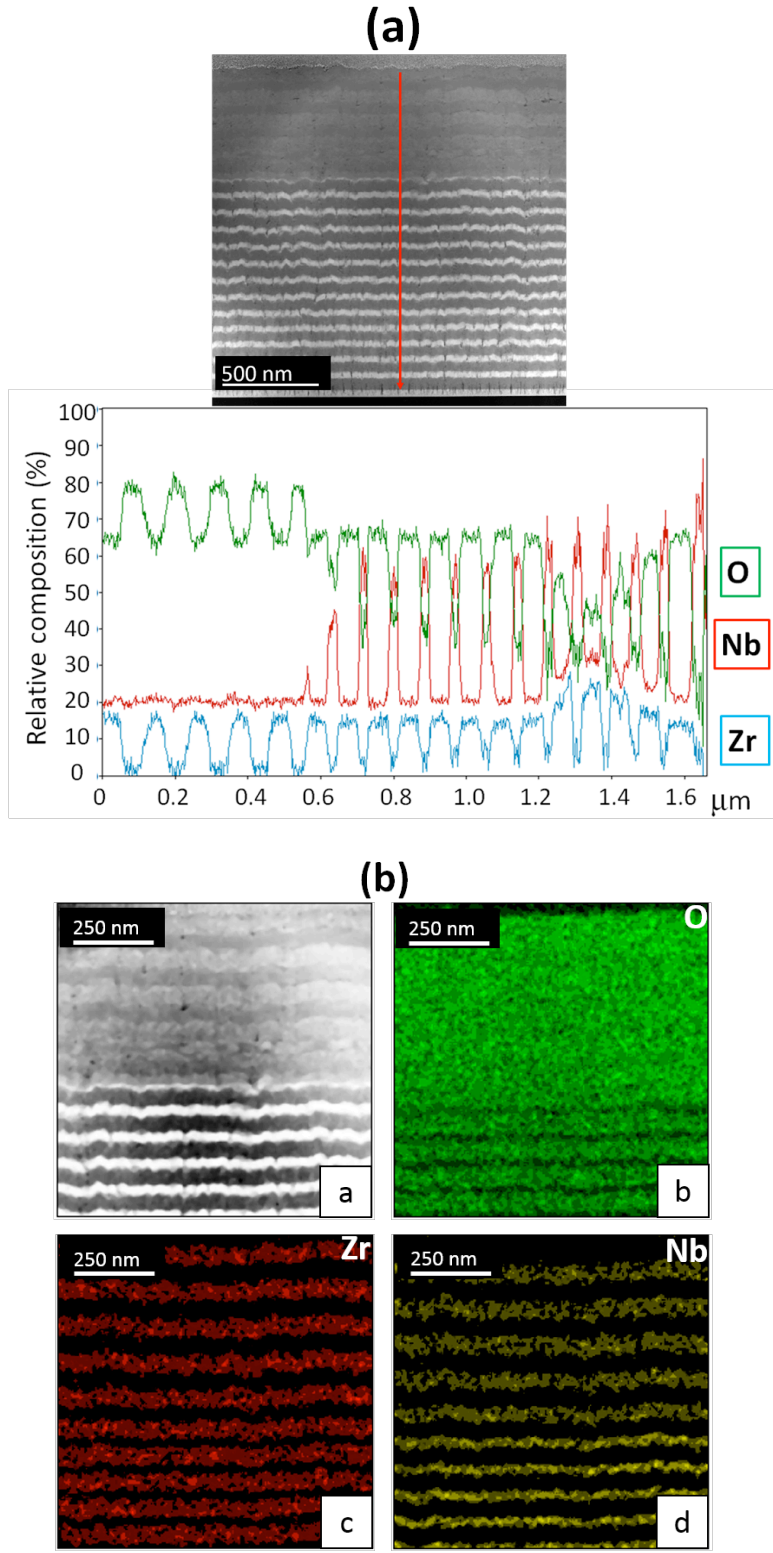

Fig. 5. (a) STEM-HAADF image of the Zr/Nb multilayer annealed for 168 h where the red arrow indicates the direction of the EELS line scan shown below and (b) STEM-EDS elemental maps of the Zr/Nb multilayer annealed for 168 h.

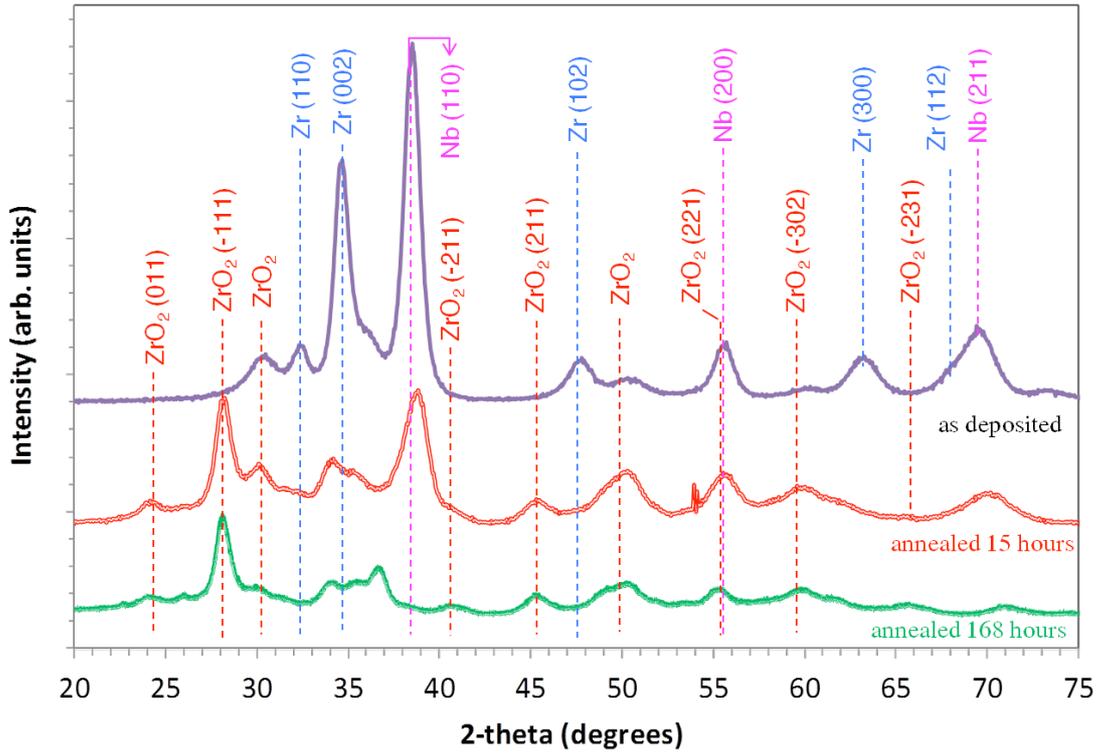

Fig. 6. XRD patterns of the as-deposited (ω = 5°) and annealed (ω = 1.5°) Zr/Nb multilayers.

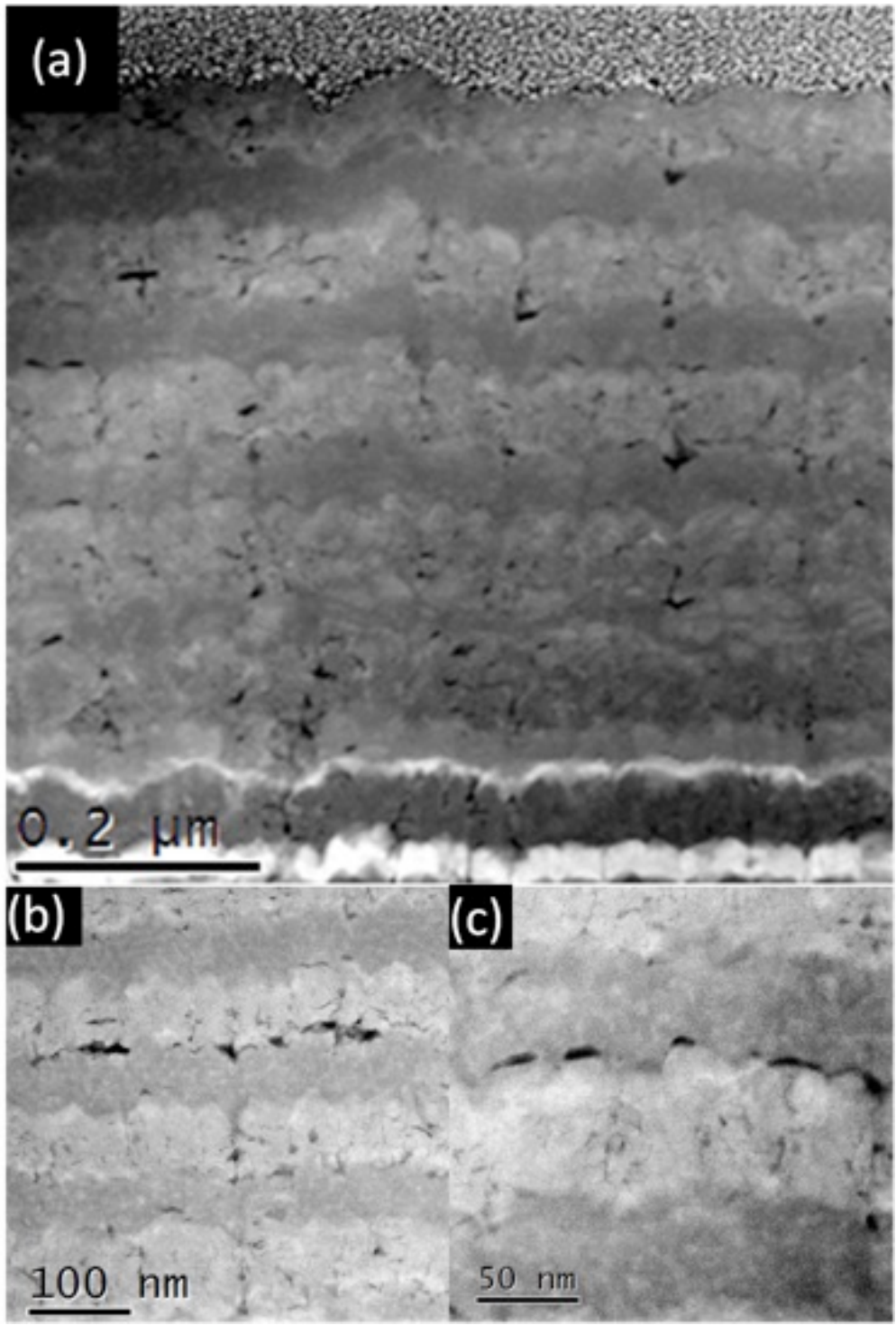

Fig. 7. (a) STEM-HAADF (JEOL ARM 200F – STEM ) images of the Zr/Nb multilayer oxidised for 168 hrs. (b) and (c) High magnification details of the fully oxidised region ($ZrO_2$ layer appear with a brighter contrast).

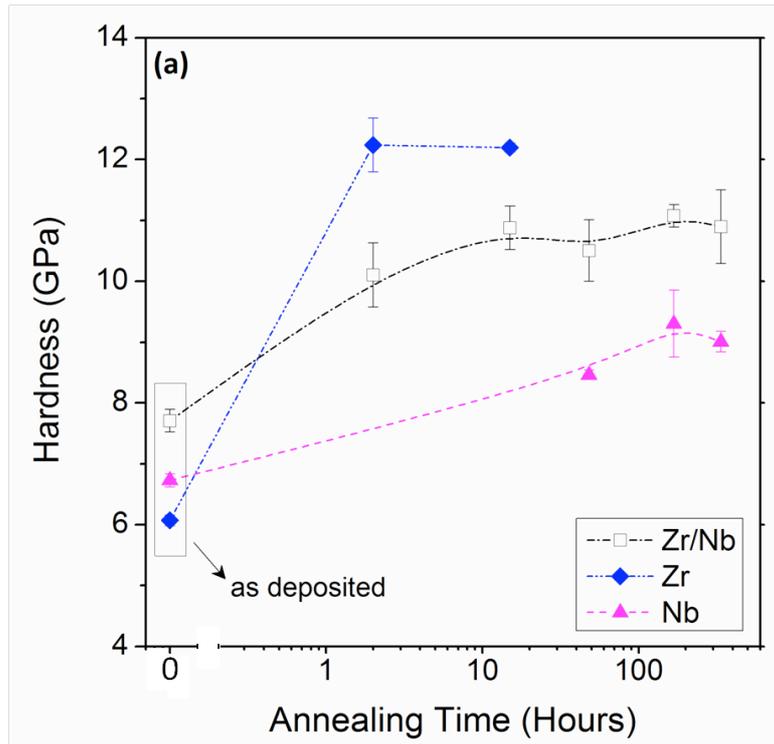

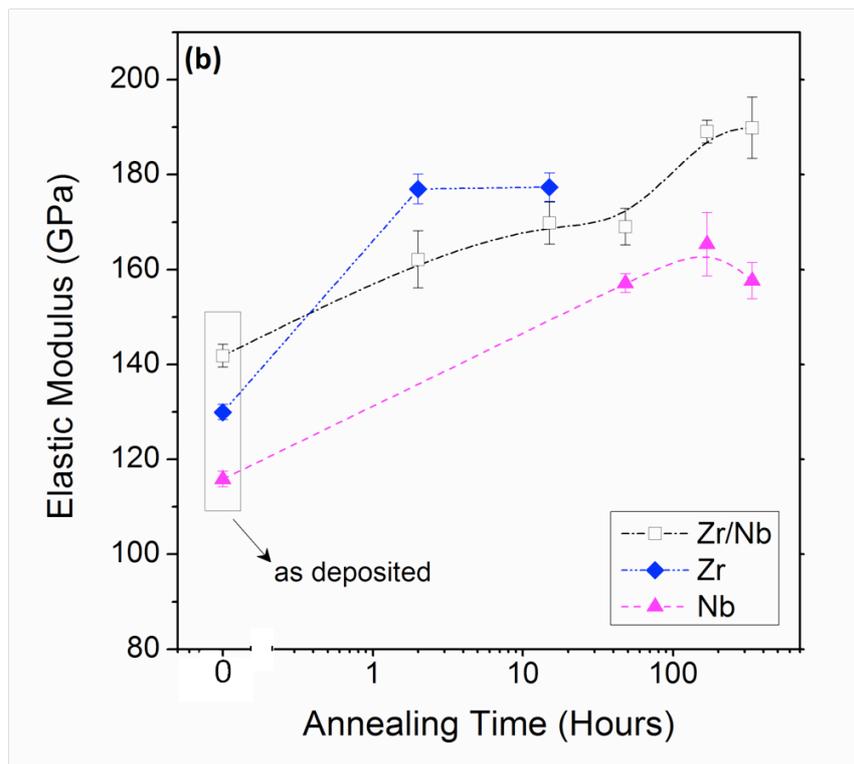

Fig. 8. (a) Hardness and (b) elastic modulus of the Zr/Nb multilayer, monolithic Zr and monolithic Nb as a function of the annealing time at 350ºC.

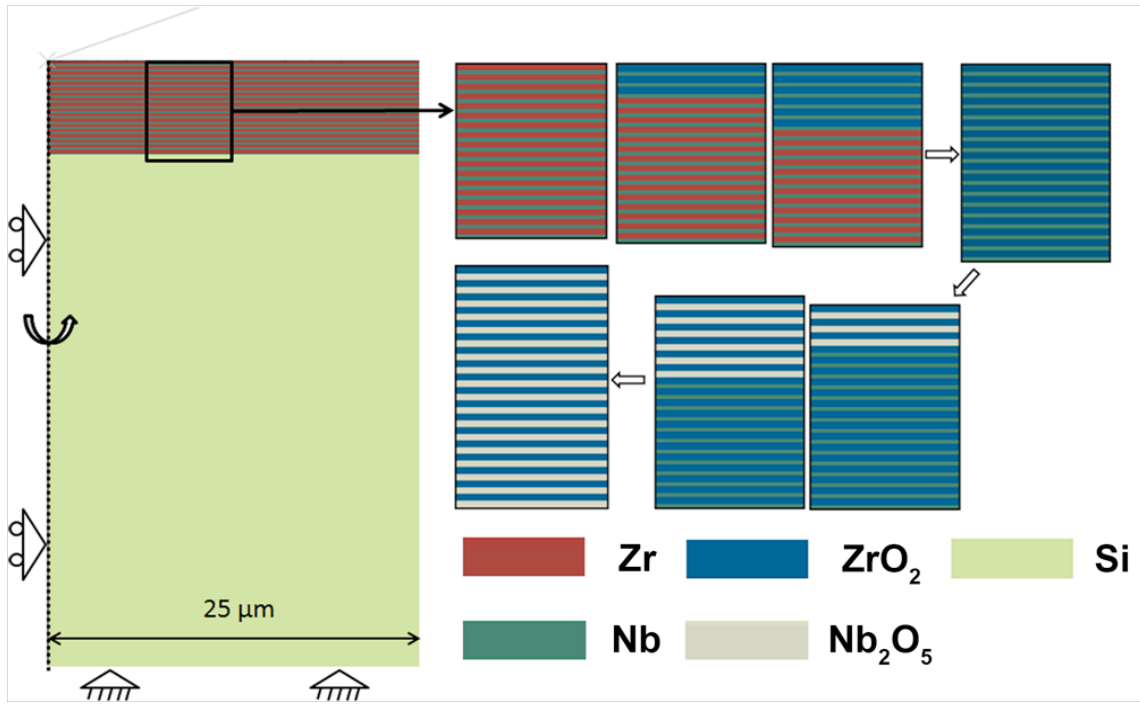

Fig. 9. Schematic of the axisymmetric finite element model of the Zr/Nb multilayer with different number of oxidised Zr and Nb layers.

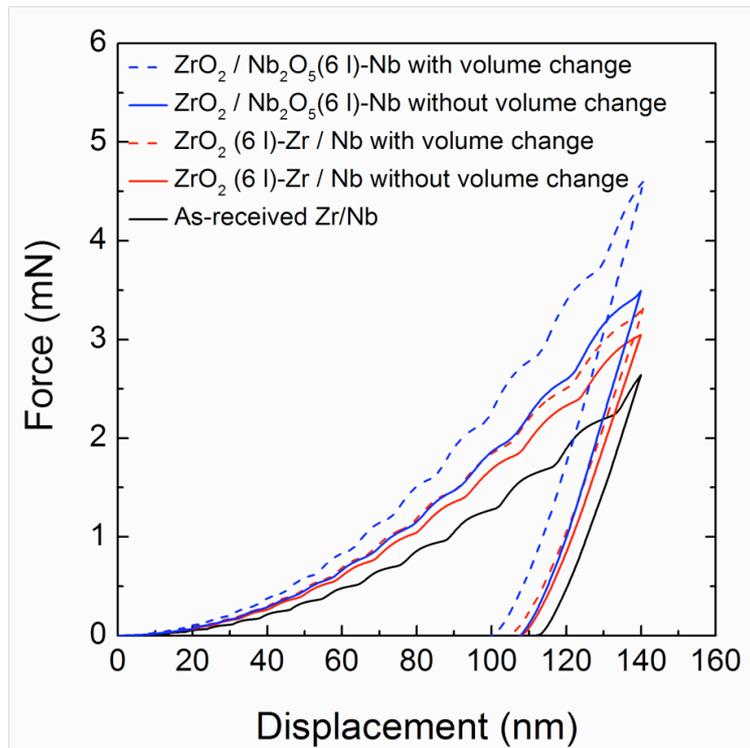

Fig. 10. Representative indentation force-displacement curves for the as-deposited Zr/Nb multilayer, ZrO$_2$-Zr/Nb multilayer with 6 ZrO$_2$ layers and ZrO$_2$/ Nb$_2$O$_5$-Nb multilayer with 6 Nb$_2$O$_5$ layers with and without accounting for internal stresses induced by the volume expansion due to oxidation.

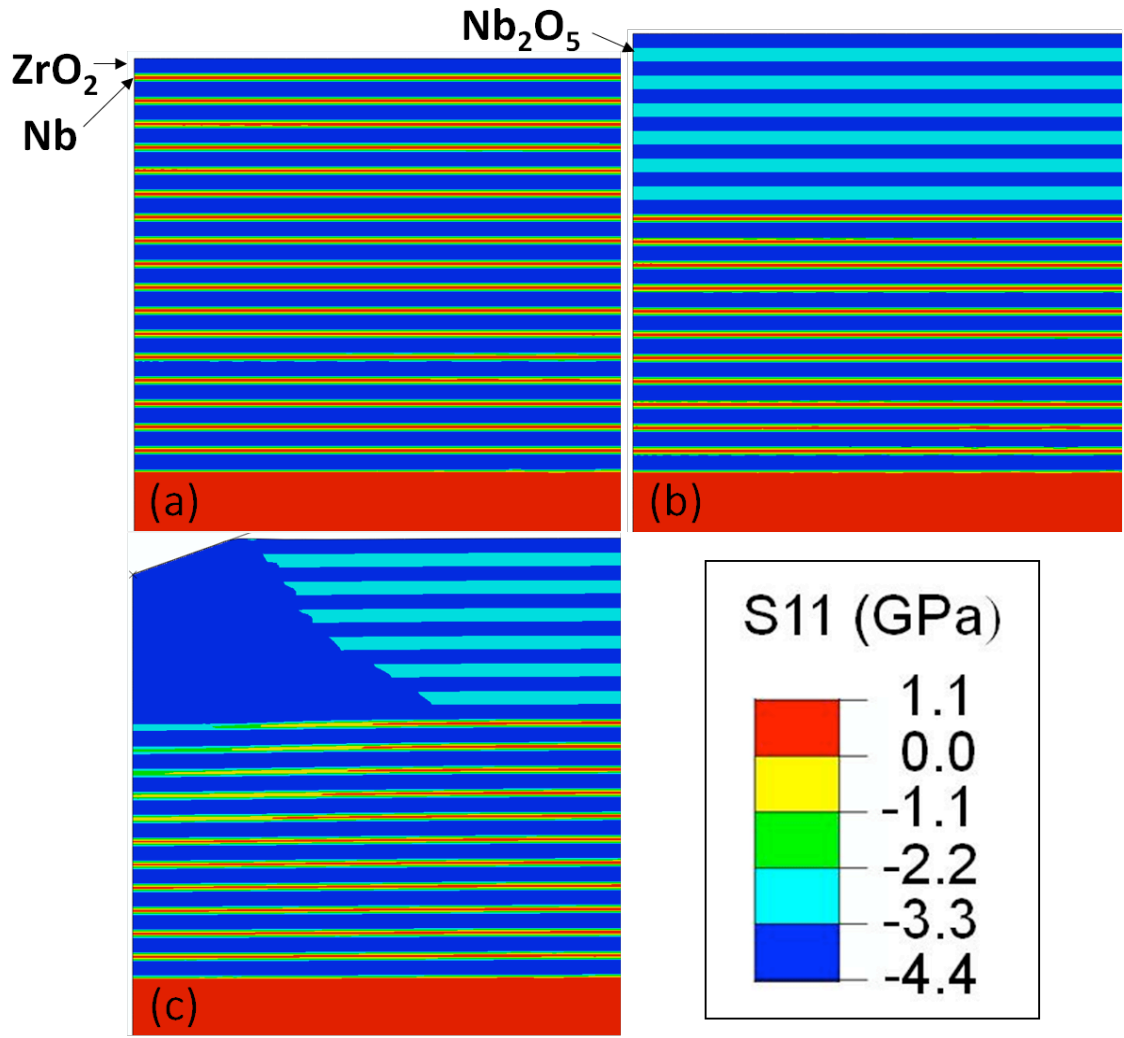

Fig. 11. Contour plot of the stresses in the radial direction: (a) Unloaded $ZrO_2$/Nb multilayer; (b) Unloaded $ZrO_2$/ $Nb_2O_5$-Nb with 6 $Nb_2O_5$ layers; (c) $ZrO_2$/ $Nb_2O_5$-Nb with 6 $Nb_2O_5$ layers loaded up to 140 mN.

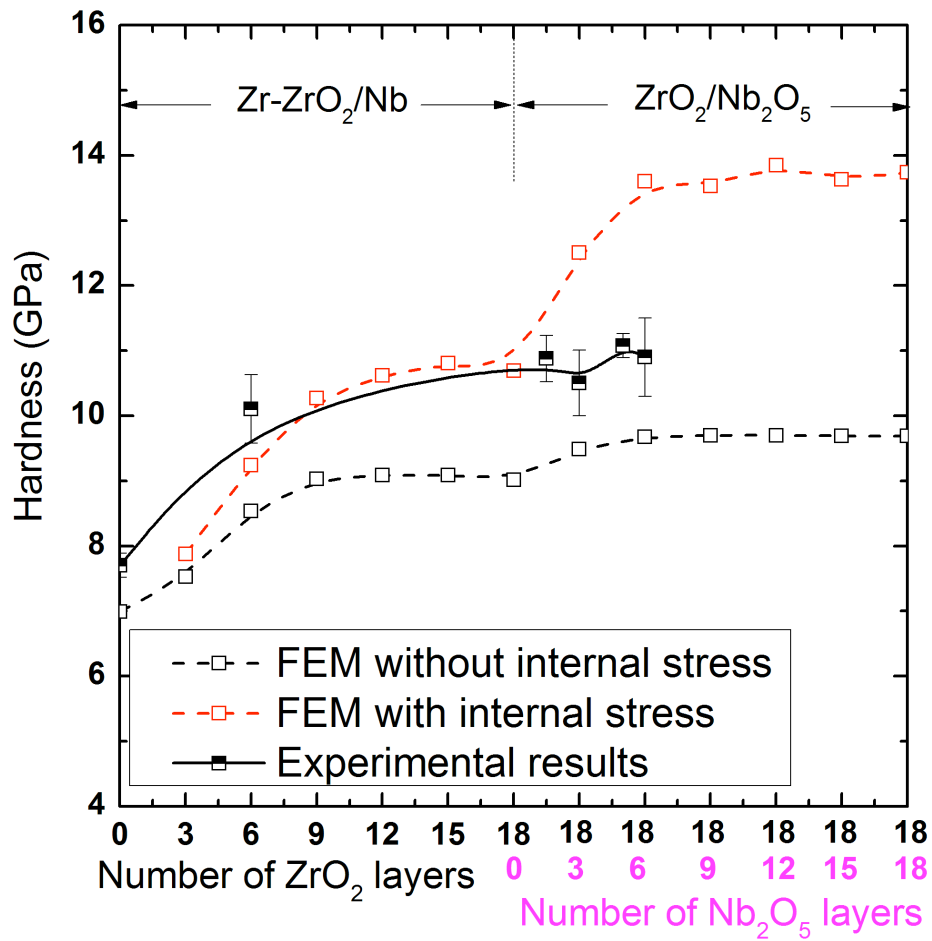

Fig. 12. Evolution of multilayer hardness as a function of the number of oxidised Zr and Nb layers with and without including the internal stresses induced by the volume change due to oxidation.

**List of tables**

Table I: Elasto-plastic properties of the materials implemented in the FE models.

| Material | Elastic Modulus (GPa) | Poisson's Ratio | Yield Stress (MPa) |
|---|---|---|---|
| Zr | 135 | 0.34 | 2228 |
| $ZrO_2$ | 175 | 0.34 | 4207 |
| Nb | 113 | 0.4 | 2467 |
| $Nb_2O_5$ | 155 | 0.4 | 3107 |
| Si | 187 | 0.28 | |